\documentstyle[prb,aps,psfig,twocolumn]{revtex}

\begin{document}
\twocolumn[\hsize\textwidth\columnwidth\hsize\csname
@twocolumnfalse\endcsname

\title{Charge ordering and chemical potential shift
 in La$_{2-x}$Sr$_x$NiO$_4$ studied by photoemission spectroscopy}
\author{M.~Satake, K. Kobayashi}
\address{Department of Physics, University of Tokyo, Bunkyo-ku,
 Tokyo 113-0033, Japan}
\author{T.~Mizokawa, A.~Fujimori\cite{adr}}
\address{Department of Physics and Department of Complexity Science and Engineering,\\
 University of Tokyo, Bunkyo-ku, Tokyo 113-0033, Japan}
\author{T.~Tanabe, T.~Katsufuji and Y.~Tokura}
\address{Department of Applied Physics, University of Tokyo,
  Bunkyo-ku, Tokyo 113-0033, Japan}
\date{\today} 
\maketitle

\begin{abstract}
We have studied the chemical potential shift in
La$_{2-x}$Sr$_x$NiO$_4$ and the charge ordering transition 
in La$_{1.67}$Sr$_{0.33}$NiO$_4$ by photoemission spectroscopy.
The result shows a large ($\sim$ 1 eV/hole) downward shift of the  
chemical potential with hole doping 
in the high-doping regime ($\delta \gtrsim$ 0.33) 
while the shift is suppressed 
in the low-doping regime ($\delta \lesssim$ 0.33).
This suppression is attributed to a segregation of doped holes on a microscopic scale
when the hole concentration is lower than $\delta \simeq 1/3$. 
In the $\delta = 1/3$ sample, 
the photoemission intensity at the chemical potential vanishes
below the charge ordering transition temperature $T_{\rm CO} =$ 240 K.
\end{abstract}

]

Recently charge ordering phenomena in transition-metal oxides 
have attracted considerable interest, 
particularly due to the possible relationship between charge stripes 
and high-temperature superconductivity
in the cuprates or giant magnetoresistance in the manganites. 
Charge ordering in a stripe form 
in the nickelates La$_{2-x}$Sr$_x$NiO$_4$ has been
well established experimentally~\cite{TranquadaNi} 
and theoretically~\cite{Zaanen} 
while it is more controversial in the cuprates, 
e.g., La$_{2-x-y}$Nd$_y$Sr$_x$CuO$_4$
with $x \simeq \frac{1}{8}$.~\cite{TranquadaCu}
In La$_{2-x}$Sr$_x$CuO$_4$, charge stripes, if exist, are dynamical, half-filled with 
holes and run along the Cu-O bond directions, whereas in La$_{2-x}$Sr$_x$NiO$_4$
they are static, filled with holes, and run along 
the diagonal Ni-Ni directions.
In order to characterize the behaviors of the stripe fluctuations 
in the cuprates and to elucidate their relationship to the superconductivity, 
it is important to study the well characterized case 
of the nickelates in more detail.
It has been established that the charge ordering 
in the diagonal stripe form in
La$_{2-x}$Sr$_x$NiO$_4$ is most stable at $x \simeq 0.33$. 
~\cite{TranquadaNi,Cheong_Resistivity,Lee,Yoshizawa}
For this composition, neutron scattering studies have shown that charge ordering
occurs below $T_{\rm CO}\simeq$ 240 K and antiferromagnetic ordering occurs
below $T_{\rm N}=$ 190 K~:\cite{Lee}
Electrical resistivity shows a steep increase below 
$T_{\rm CO}$.\cite{Cheong_Resistivity}

In this paper, we report on a photoemission study 
of the chemical potential shift in La$_{2-x}$Sr$_x$NiO$_4$ 
as a function of hole concentration and the temperature dependence
of photoemission spectra across the phase transitions 
for $x=0.33$.
The shift of the electron chemical potential $\mu$ 
as a function of electron density $n$
is related to the charge compressibility $\kappa$ 
or the charge susceptibility $\chi _{\rm c}$ through 
$\kappa = (1/n^2)(\partial n/\partial \mu)$ or 
$\chi _{\rm c}= \partial n/\partial \mu$ and can be measured 
through the shifts of spectral features in photoemission spectra 
since binding energies in the photoemission spectra 
are experimentally referenced to the chemical potential $\mu$, 
namely, the Fermi level.
The chemical potential shift has been studied for La$_{2-x}$Sr$_x$CuO$_4$ 
(LSCO) and found to be suppressed in the underdoped region 
$x \lesssim 0.15$.\cite{Ino_Chemical}
The suppressed chemical potential shift, or equivalently the enhanced charge 
compressibility, in spite of the reduced density of states 
at the chemical potential as measured 
by the electronic specific heat, \cite{Oda}
has been interpreted as due to the opening of a pseudogap or 
to a microscopic phase separation.\cite{Ino_Chemical} 
Indeed, such a pseudogap opening has been confirmed by a subsequent
angle-resolved photoemission study of LSCO.\cite{Ino_DOS} 

Single crystals of La$_{2-x}$Sr$_x$NiO$_{4+y/2}$ (LSNO) were prepared
by the floating zone method.
Hole concentration $\delta = x + y$ except for the $x =$ 0 sample
was determined by iodimetric titration with an accuracy of $\pm$0.01.
The chemical compositions thus determined are 
tabulated in Table~\ref{tab0}.
X-ray photoemission spectroscopy (XPS) measurements
were performed using a spectrometer equipped with
a Mg $K\alpha$ source ($h\nu =$ 1253.6 eV)
and a PHI double-pass cylindrical-mirror analyzer.
The energy resolution including the x-ray source and the 
analyzer was $\sim$ 1.0 eV but we could determine the shifts of core levels 
to an accuracy of $\pm$50 meV as in the case of LSCO.\cite{Ino_Chemical}
Binding energies were calibrated using Au evaporated on the samples. 
All the XPS spectra were taken 
at liquid-nitrogen temperature ($\sim$ 77 K).
High-resolution ultraviolet photoemission spectroscopy (UPS)
measurements were carried out using
a He {\footnotesize I} resonance line ($h\nu=$ 21.2 eV) 
and VSW and Omicron 125-EA hemispherical analyzers.
The He {\footnotesize I} spectra have been corrected 
for the He {\footnotesize I}$^{*}$ satellite.
In order to determine the Fermi level ($E_F$) and to
estimate the instrumental resolution, 
Au was evaporated on each sample.
The energy resolution was estimated to be 25-30 meV.
The sample surfaces were repeatedly scraped {\it in situ} 
with a diamond file to obtain clean surfaces.
The cleanliness of the surfaces was checked by lack of 
contamination/degradation-related features on the higher 
binding-energy side of the O 1$s$ peak in the XPS spectra or
that at $\sim -4.5$ and $\sim -9$ eV in the UPS spectra.

\begin{figure}[p]
\begin{center}
\psfig{figure=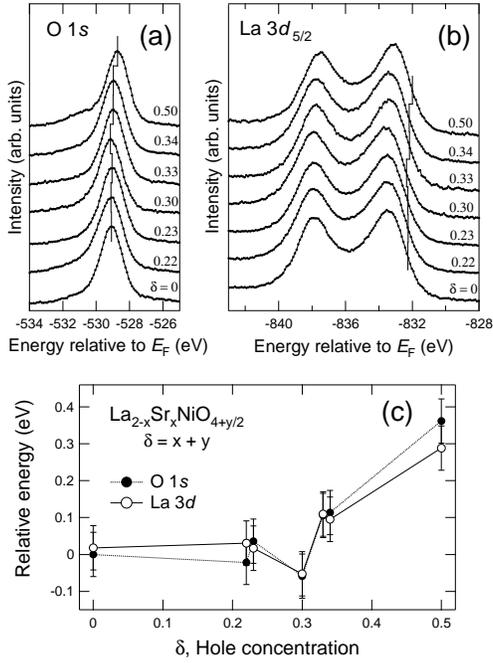,width=7cm}
\caption{O $1s$ (a) and La $3d_{5/2} (b)$ core-level spectra
 of La$_{2-x}$Sr$_x$NiO$_{4+y/2}$ taken at  $h\nu =$ 1253.6 eV.
 The hole concentration is given by $\delta = x + y$.
 Vertical lines indicate the energy shift of the spectra (see text).
 (c) Energy shifts of the  O $1s$ and La $3d$ spectra as
 functions of hole concentration.}
\label{fig:XPS}
\end{center}
\end{figure}

Figure~\ref{fig:XPS} shows the XPS spectra of the O $1s$ and
La $3d_{5/2}$ core levels taken at liquid-nitrogen temperature. 
The vertical lines 
indicate the estimated positions of the core levels.
As for the La $3d_{5/2}$ peak, the shift was estimated from 
the position at half peak height on the lower-binding-energy side
of the peak because the effect of surface degradation appears 
on the higher-binding-energy side.
Estimated shifts of the O $1s$ and La $3d$ spectra are plotted in 
Fig.~\ref{fig:XPS}(c).
Now, we can assume that the shifts of the O $1s$ and La $3d$ core levels
are largely due to the shift of the chemical potential as
in the case of LSCO\cite{Ino_Chemical}
for the following reasons.
First, the identical shifts of the O $1s$ and La $3d$ spectra
with $\delta$ indicate that the effect of changes in the Madelung potential 
caused by the La$^{3+} \rightarrow$ Sr$^{2+}$ substitution 
can be neglected. This is because the changes in the Madelung potential 
would cause shifts the core levels of the O$^{2-}$ anion
and the La$^{3+}$ cation in the opposite directions if this effect were significant. 
Second, changes in the number of electrons of the 
O and La atoms with hole concentration, which may cause shifts of 
the La and O core levels, can also be neglected. 
Unfortunately, it was not possible to measure the shift of the Ni $2p$ core level 
with sufficient accuracy because the Ni $2p_{3/2}$ peak 
was overlaid by the La $3d_{3/2}$ peak and 
the Ni $2p_{1/2}$ peak was too broad.

\begin{figure}[p]
\begin{center}
\psfig{figure=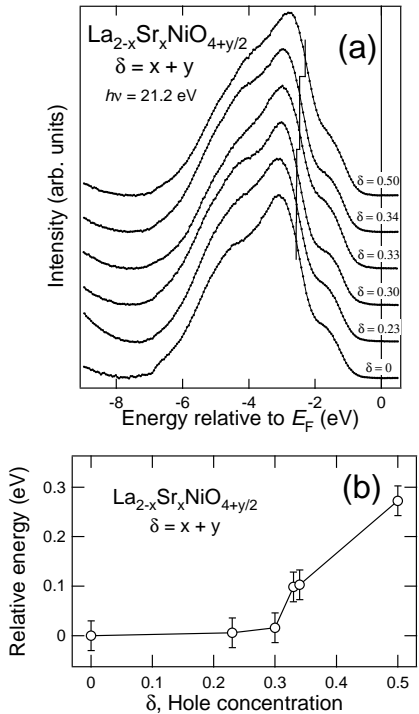,width=6cm}
\caption{(a)Valence-band spectra of La$_{2-x}$Sr$_x$NiO$_{4+y/2}$.
  Vertical lines indicate the position at 3/4 height of the non-bonding 
  O $2p$ peak on the lower binding-energy side.
  (b)Energy shift of the non-bonding O $2p$ peak in the valence-band
  spectra as a function of hole concentration.}
\label{fig:UPS}
\end{center}
\end{figure}

\begin{figure}[p]
 \begin{center}
\psfig{figure=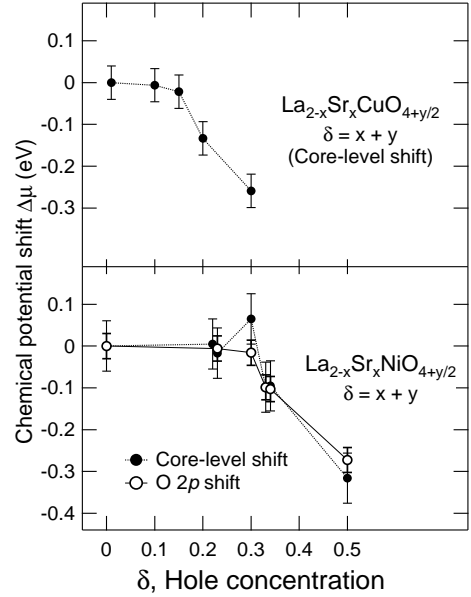,width=7cm}
  \caption{(a) Chemical potential shift $\Delta\mu$
 in La$_{2-x}$Sr$_x$NiO$_{4+y/2}$ as a function of hole
 concentration $\delta$.
  Solid circles are $\Delta\mu$
  deduced from the average shift of the O $1s$ and La $3d$ XPS.
  Open circles indicate $\Delta\mu$ deduced from the shift of 
  the non-bonding O $2p$ peak in the valence-band UPS. 
  The UPS data for $\delta =$ 0 was taken at 230 K,
  and the others at 150 K.
  (b) Chemical potential shift $\Delta\mu$ in La$_{2-x}$Sr$_x$CuO$_4$ 
  as a function of hole concentration $\delta$
  (taken from Ref.~\protect\onlinecite{Ino_Chemical}). }
  \label{fig:ChemicalPS}
 \end{center}
\end{figure}

The valence-band spectra of LSNO with
various hole concentrations $\delta = x + y$ were also measured
using UPS as shown in Fig.~\ref{fig:UPS}(a) and
the spectral shift was deduced in the same way.
All the spectra were taken at 150 K except for $\delta = 0$,
whose spectrum was measured at 230 K to avoid charging effect.
These spectra have been normalized to the peak height at $\sim -3$ eV.
The shift was estimated from the non-bonding O $2p$ peak
at $\sim -3$ eV because its electronic state 
should be insensitive to the hole concentration
compared to the $d^8\underline{L}$ final states at $\sim -1.5$ eV.
In order to avoid the effect of surface degradation,
which appear at $\sim-4.5$ eV, the shift of the O $2p$ peak was estimated from 
the position at 3/4 peak height on the lower-binding-energy side 
of the peak as indicated by vertical lines in Fig.~\ref{fig:UPS}(a).
Figure~\ref{fig:UPS}(b) shows the shift of the O $2p$ peak thus evaluated.
The shift of the non-bonding O $2p$ peak should represents the chemical
potential shift $\Delta\mu$ in the same way as that of the O $1s$ peak.
Indeed, $\Delta\mu$ deduced from the
shift of the O $2p$ levels and that deduced from 
the average shift of the O $1s$ and La $3d$ levels 
agree with each other as shown in Fig.~\ref{fig:ChemicalPS}(a),
where $\Delta E_{{\rm O} 2p}$, $\Delta E_{{\rm O} 1s}$,
and $\Delta E_{{\rm La} 3d}$ are changes 
in the binding energies of the O $2p$, O $1s$, and La $3d$ levels, 
respectively.

The sign of $\Delta\mu$ is consistent with the downward shift
as expected for hole doping.
However, the suppression of the shift for $0 \le \delta \le 0.30$
cannot be explained within a rigid-band picture
because the chemical potential $\mu$ in an insulator 
would shift rapidly with hole doping in the rigid-band picture. 
Therefore, this non-rigid-band behavior implies 
remarkable correlation effects in LSNO. 
Such a suppression of $\Delta\mu$ has also been
found in LSCO as shown in Fig.~\ref{fig:ChemicalPS}(b),\cite{Ino_Chemical} 
where $\mu$ shows a large ($\sim$ 1.5 eV/hole) downward shift with hole 
doping in the overdoped ($x \gtrsim$ 0.15) region 
and a small ($<$ 0.2 eV/hole) shift 
in the underdoped ($x \lesssim$ 0.15) region. 
In LSNO, $\mu$ shows a large ($\sim$ 1.5 eV/hole) downward shift
for $\delta \gtrsim$ 0.33 and a small ($<$ 0.2 eV/hole) shift for 
$\delta \lesssim$ 0.33. The low-doping region of $x \lesssim$ 0.33 
in the nickelates is thus analogous to the underdoped region
of $x \lesssim$ 0.15 in the cuprates.
Since LSCO is considered to show dynamical stripe fluctuations
according to the inelastic neutron scattering 
studies,\cite{Yamada_Neutron} it is likely that 
the pinning of $\mu$ below the critical hole concentration 
is a common phenomenon for stripe formation.
The following scenario may be considered as a common mechanism for the 
pinning of $\mu$.
Charge ordering in a stripe form is a kind of ``phase separation'' 
into a hole-rich region
and a hole-poor region on a microscopic length scale.
The hole chemical potential $-\mu$ (where $\mu$ is the electron chemical 
potential) 
is the energy required to add one hole to the system.
In general, increased holes concentration increases the average 
hole-hole repulsion per hole 
and hence increases $-\mu$ if the hole distribution 
is spatially uniform.
Therefore, the absence of change in $-\mu$ with hole doping suggests
that there is no increase in the average hole-hole repulsion with 
increasing hole concentration.
This can be made possible for a system in which holes are segregated, 
e.g., in a stripe form.
For $\delta \gtrsim 0.33$ there is large downward shift of 
$\mu$ with hole doping, indicating an increase
in the hole-hole repulsion between the overdoped holes.
Therefore, it is considered that the repulsive interaction 
between hole stripes becomes
significant for  $\delta > 1/3$, where the stable 
$\epsilon = 1/3$ stripe ordering cannot survive.
It is interesting to note that the shift of $\mu$ for LSCO and LSNO
in the overdoped region is the same in magnitude in spite of
their quite different electronic properties, i.e., metallic {\it versus} 
insulating.

A small but finite jump of $\sim0.1$ eV $\mu$ from $\delta =$ 0.30 
to $\delta =$ 0.33 appears to exist, according to Fig.~\ref{fig:ChemicalPS}.
This jump may be explained by the gap opening in the
$\delta =$ 0.33 material as observed by the optical
conductivity measurements:\cite{Katsufuji_Optical}
If the gap remains stable for a finite range of hole concentration 
around $\delta = 1/3$ due to the high stability 
of the $\epsilon = 1/3$ stripe,
in going from $\delta\lesssim 1/3$ to $\delta\gtrsim 1/3$, 
$\mu$ should show a jump equal to the magnitude of gap. 
The jump of $\sim$0.1 eV is smaller than the gap value of 0.26 eV
estimated from the optical spectra.\cite{Katsufuji_Optical}
The discrepancy can be explained by the different experimental 
methods because the optical measurements probe the
direct (momentum-conserving) gap whereas the chemical potential shift 
probes the minimum gap (which may be indirect).

\begin{figure}[p]
\begin{center}
\psfig{figure=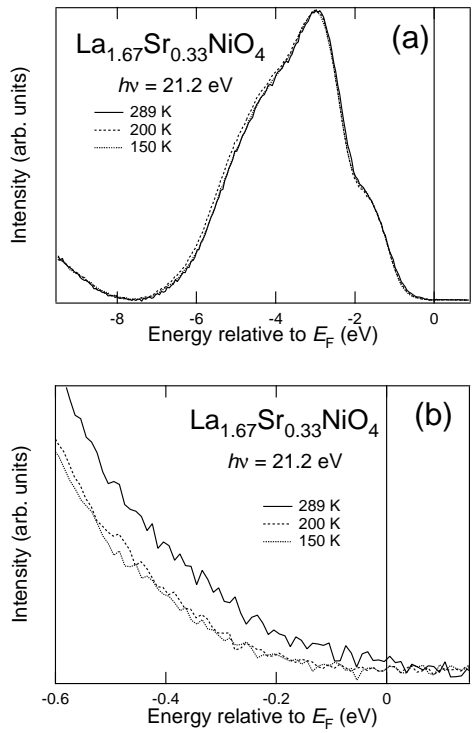,width=7cm}
\caption{Valence-band spectra of La$_{1.67}$Sr$_{0.33}$NiO$_4$
  taken at 150, 200, and 289 K in a wide energy
  range (a) and near the chemical potential (b).}
\label{fig:DOS_Temp}
\end{center}
\end{figure}

Figure~\ref{fig:DOS_Temp} shows valence-band spectra of
 La$_{1.67}$Sr$_{0.33}$NiO$_4$ taken at 150, 200, and 289 K 
in order to study spectral changes above and below the
charge-ordering temperature $T_{\rm CO}$ = 240 K.
Here, the background has been subtracted assuming the
secondary-electron cascade process\cite{Li_Background} and the
spectra have been normalized to the peak height at $\sim -3$ eV.
It should be noted that the following interpretation of the spectra is
 almost independent of the normalization procedure.

Each spectrum near $E_F$ in Fig.~\ref{fig:DOS_Temp}(b) shows
no Fermi edge, which is consistent with the insulating behavior
below room temperature.\cite{Cheong_Resistivity}
The intensity near $E_F$ clearly changes between above 
(289 K) and below (150 and 200 K) $T_{\rm CO}$ 
although there are no significant changes 
in the wide range spectra shown in Fig.~\ref{fig:DOS_Temp}(a).
It should be noted that the $d^8\underline{L}$
feature at $\sim -1.5$ eV is significantly broadened above $T_{\rm CO}$.
This broadening is much stronger than that expected 
>from the temperature broadening between
289 and 200 K and seems to reflect the order-to-disorder 
transition of charge carriers at $T = T_{\rm CO}$.
The decrease of the intensity at $E_{\rm F}$ in going from 289 
to 200 K would be due to the gap opening 
at $T_{\rm CO}$ caused by the charge ordering
as observed in the optical conductivity.\cite{Katsufuji_Optical}
On the other hand, there is little change in the intensity between
150 and 200 K, indicating that the spectral intensity is not much affected 
by the spin ordering across $T_N$ (= 190 K).
These results are consistent with the electrical 
resistivity,\cite{Katsufuji_Optical} which shows a large change 
between 200 and 289 K and little changes between 150 and 200 K.

In conclusion, we have observed that 
the chemical potential in La$_{2-x}$Sr$_x$NiO$_4$
shows a large ($\sim$ 1 eV/hole) downward shift with hole
concentration in the high doping regime ($\delta \gtrsim$ 0.33) while 
it shows no appreciable shift in the
low doping regime ($\delta \lesssim$ 0.33).
We have explained this observation as due to a segregation of doped holes 
on a microscopic length scale
when the hole concentration is lower than $\delta = 1/3$, where the stable 
$\epsilon = 1/3$ charge stripes are formed.
In La$_{1.67}$Sr$_{0.33}$NiO$_4$, the photoemission spectra exhibit subtle changes 
across the charge ordering transition temperature $T_{\rm CO} =$ 240 K: 
the intensity at $E_{\rm F}$ vanishes below  $T_{\rm CO}$, consistent 
with the transport and optical properties. 

The authors would like to thank Dr. Y. Aiura 
for useful advise in designing the sample holder.
MS is indebted to H. Hayashi for collaboration in the initial stage of 
this work. 
This work was supported by a Special Coordination Fund from the 
Science and Technology of Japan and Agency and the New Energy and Industrial
Development Organization (NEDO).


\begin{table}
\caption{Chemical compositions of the La$_{2-x}$Sr$_x$NiO$_{4+y/2}$
samples studied in the present work. $\delta = x + y$ gives the hole 
concentration. No iodimetric titration was made for the $x=0$ sample.}
\label{tab0}
\begin{tabular}{crrrrrrr}
$\delta$&0&0.22&0.23&0.30&0.33&0.34&0.50\\
\tableline
$x$&0&0.10&0.20&0.30&0.33&0.36&0.50\\
$y$&0&0.12&0.03&0.00&0.00&-0.02&0.00
\end{tabular}
\end{table}

\begin{references}
\bibitem[*]{adr}To whom correspondence should be sent.
\end{references}
\end{document}